\documentclass[prx,aps,twocolumn,english,10pt,superscriptaddress,floatfix,longbibliography]{revtex4-2}
\usepackage{multibib}
\usepackage{booktabs}
\usepackage{times}
\usepackage{braket}
\usepackage{mathptmx}
\usepackage{cancel}
\usepackage{adjustbox}
\usepackage{amsmath, amsthm, thm-restate, bbold, physics, amssymb}
\usepackage{braket, enumitem, relsize}
\usepackage{makecell, graphicx, subfiles, multirow, wrapfig}

\usepackage[font=scriptsize]{subcaption}
\usepackage[font=scriptsize, justification=raggedright]{caption}
\usepackage{subcaption}
\usepackage[toc,page]{appendix}
\usepackage[colorlinks, allcolors=black]{hyperref}
\usepackage[capitalize, nameinlink]{cleveref}
\usepackage{xcolor, tabularx}
\setlist[itemize]{leftmargin=10pt}
\usepackage[utf8]{inputenc}
\usepackage{float}
\usepackage{tikz}
\usetikzlibrary{quantikz}
\usepackage{xcolor}
\hypersetup{
  colorlinks   = true, 
  urlcolor     = teal, 
  linkcolor    = teal, 
  citecolor   = teal 
}

\mathcode`l="8000
\begingroup
\makeatletter
\lccode`\~=`\l
\DeclareMathSymbol{\lsb@l}{\mathalpha}{letters}{`l}
\lowercase{\gdef~{\ifnum\the\mathgroup=\m@ne \ell \else \lsb@l \fi}}%
\endgroup

\renewcommand{\epsilon}{\varepsilon}

\newtheorem*{thm*}{Theorem}

\makeatletter
\newcounter{savesection}
\newcounter{apdxsection}
\renewcommand\appendix{\par
  \setcounter{savesection}{\value{section}}%
  \setcounter{section}{\value{apdxsection}}%
  \setcounter{subsection}{0}%
  \gdef\thesection{\@Alph\c@section}}
\newcommand\unappendix{\par
  \setcounter{apdxsection}{\value{section}}%
  \setcounter{section}{\value{savesection}}%
  \setcounter{subsection}{0}%
  \gdef\thesection{\@arabic\c@section}}
\makeatother

\begin{abstract}
Our study aims to increase the spatial resolution of high-sensitivity magnetometry based on singlet-transition infrared (IR) absorption using nitrogen-vacancy (NV) centers in diamonds in monolithic cavities, with potential applications in bio-magnetic field detection. We develop a master-equation treatment of optically detected magnetic resonance, incorporating IR light saturation effects. This master equation provides the singlet population, which is then utilized to calculate the reflectivity and ultimately derive the magnetic field sensitivity taking into account photon and spin shot noise. We further show that our model is compatible with experiments of IR-based NV center magnetometry. Through optimization in a high-parameter space, we uncover the potential to achieve sensitivities in the order of sub-pico tesla, even for sub-millimeter scales.

\end{abstract}

\begin{document}

\title{Master equation-based model for infrared-based magnetometry with nitrogen-vacancy centers in diamond cavities: a path to sub-picotesla sensitivity at sub-millimeter scales}

\author{Hadi Zadeh-Haghighi}
\email{hadi.zadehhaghighi@ucalgary.ca}
\affiliation{Department of Physics and Astronomy, University of Calgary, Calgary, AB T2N 1N4, Canada}
\affiliation{Institute for Quantum Science and Technology, University of Calgary, Calgary, AB T2N 1N4, Canada}

\author{Omid Golami}
\affiliation{Department of Physics and Astronomy, University of Calgary, Calgary, AB T2N 1N4, Canada}
\affiliation{Institute for Quantum Science and Technology, University of Calgary, Calgary, AB T2N 1N4, Canada}

\author{Vinaya Kumar Kavatamane}
\affiliation{Department of Physics and Astronomy, University of Calgary, Calgary, AB T2N 1N4, Canada}
\affiliation{Institute for Quantum Science and Technology, University of Calgary, Calgary, AB T2N 1N4, Canada}

\author{Paul E. Barclay}
\affiliation{Department of Physics and Astronomy, University of Calgary, Calgary, AB T2N 1N4, Canada}
\affiliation{Institute for Quantum Science and Technology, University of Calgary, Calgary, AB T2N 1N4, Canada}
\affiliation{Nanotechnology Research Centre, National Research Council of Canada, Edmonton, AB T6G 2M9, Canada}

\author{Christoph Simon}
\email{csimo@ucalgary.ca}
\affiliation{Department of Physics and Astronomy, University of Calgary, Calgary, AB T2N 1N4, Canada}
\affiliation{Institute for Quantum Science and Technology, University of Calgary, Calgary, AB T2N 1N4, Canada}

\date{\today}

\maketitle

\section{Introduction}
Detecting minuscule magnetic fields plays a pivotal role in various scientific and technological domains, including geoscience, atmospheric science, atomic and particle physics, (bio)chemistry, and medicine \cite{Degen_2017,Ripka2010}. Quantum sensors, harnessing inherent quantum properties, offer extraordinary precision, repeatability, and accuracy in measurements \cite{Budker_2007}. These sensors find applications across diverse technologies \cite{Fu2020}. While current sensors are already highly sensitive, ongoing technological advancements aim to meet the growing demands in various domains.  For instance, applications like magnetocardiography and magnetoencephalography continue to benefit from improvements in both sensitivity and spatial resolution \cite{Arai_2022, Baillet_2017}. Wearable and highly sensitive magnetic sensors with superior spatial resolution would substantially aid in medical diagnostics.

\par
Such ranges of magneto-sensing have been achieved based on technologies such as optically pumped magnetometers \cite{Kim_2016,Zhang_2022}, superconducting quantum interference devices (SQUIDs) \cite{Storm_2017}, and tunnelling magnetoresistance sensors \cite{Gao_2020}. 
However, these magnetometers typically offer spatial resolutions that are confined to a few centimetres due to limitations in standoff distance or sensor size. Notably, SQUIDs, with their requirement for cryogenic conditions and extended sensor geometries, operate with standoff distances of around a centimetre or more from the biological sample. It is worth mentioning that magnetic resonance force microscopy (MRFM) is an imaging technique that can achieve angstrom spatial resolution but requires non-ambient conditions such as high vacuum \cite{Rugar2004}. Thus, the next crucial step in advancing magnetometers is to enhance their spatial resolution to sub-centimetre scales. 
\par

\begin{figure*}
    \centering
 \includegraphics[width=0.9\textwidth]{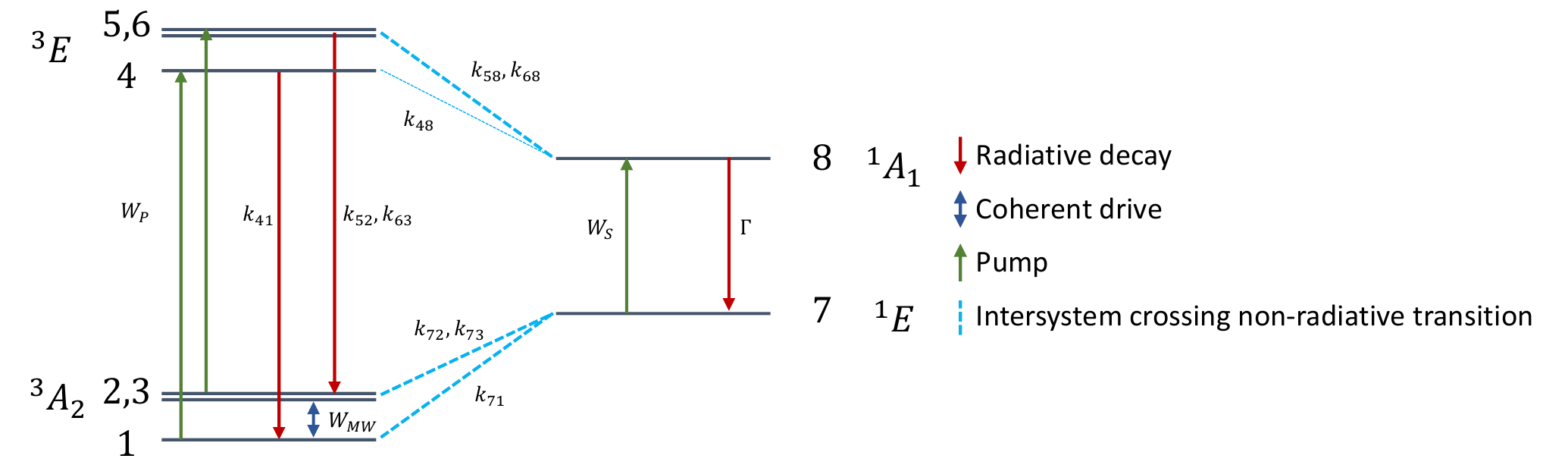}
    \caption{The electronic structure of the NV$^-$ color center in diamond at zero magnetic field. The $^3A_2$ and $^3E$ states are triplet states, while $^1E$ and $^1A_1$ are singlet states. The intersystem crossing (ISC) from the spin $m_\text{s} = 0$ sub-level of the excited triplet state to the singlet state is weaker compared to the ISC from the spin $m_\text{s} = \pm1$ sub-levels. labels 1-8 represent the energy levels employed in the master equation analysis. The transition rates for different level are shown by $\kappa$ and $\Gamma$, and values given in Table \ref{tab:Table1} in Appendix B.}
        \label{fig:Figure1}
\end{figure*}

Among recent sensors, the use of nitrogen-vacancy (NV) color centers in diamond for magnetometry has gained considerable attention \cite{Schirhagl_2014,Kucsko_2013,Lovchinsky_2016,Barry2016,Wu_2016, Glenn_2017, Casola_2018,Shi_2018,Liu_2019,Liu_2020,Li2021,Wolfowicz2021,Soshenko2021,Zhang2021,Hahl2022,Shim2022,Dwyer2022,Wang2022,Barry_2020}, as they function under ambient conditions. In the negatively charged NV center in diamond, the transition between the ground and excited triplet states is optically allowed (see Fig. \ref{fig:Figure1}) \cite{doherty2011negatively}. At room temperature and above, the electron spin resonance of the NV center can be easily observed due to the millisecond-order spin-lattice relaxation time and the fluorescence contrast between the $m_s=0$ and $m_s=\pm 1$ spin sub-levels. Resonant lasers can pump electrons from the ground state to the excited triplet state while conserving their spin. These excited electrons can decay either by emitting a photon or through intersystem crossing to the $^1A_1$ singlet state, which is non-radiative and non-spin-conserving. From there, the electron decays to the $^1E$ metastable state by emitting an IR (1042 nm) photon, and then non-radiatively to the ground triplet state. The decay from the spin $\pm1$ sub-levels to the singlet state is more pronounced than from the spin $0$ sub-level, resulting in a higher probability of non-radiative decay for electrons in the spin $\pm1$ sub-levels. Consequently, a green laser will eventually populate the spin-zero sub-level \cite{kehayias_2018}. When microwave radiation resonates with the ground state spin sub-levels, it induces transitions between these sub-levels, reducing fluorescence intensity and causing a dip in the photo-luminescence spectrum. This Optically Detected Magnetic Resonance (ODMR) signal, sensitive to external magnetic fields and temperature, is crucial for sensing applications \cite{Siyushev_2013,Doherty2013}.

\par
Aiming for better resolution in magnetic field detection, recent investigations into microcavity-enhanced NV center magnetometry have shown promise \cite{Kim2023}. This method combines the unique properties of NV centers in diamond with enhanced light-matter interaction within microcavities to achieve highly sensitive magnetic field measurements and potential spatial resolution improvements \cite{Vahala_2003,Maze2008,Aspelmeyer_2014,Faraon2011,Tisler2009,Barclay2009,Santori2010}.\\

Conventional fluorescence-based magnetometer sensitivity is constrained by shot noise due to low fluorescence contrast \cite{Wang2015,Xu2019,Zhang2021}. One approach to enhance the fidelity of readout in NV-center magnetometry is the readout based on the population in the NV singlet states \cite{Jensen2014}. At room temperature, the upper singlet state, $^1 A_1$ (See Fig. \ref{fig:Figure1}), has a lifetime of $\lesssim 1$ ns \cite{Acosta_2010,Ulbricht2018} while the lifetime of the lower singlet state, $^1 E_1$, is $\sim$ 140-220 ns \cite{Acosta_2010,Robledo2011,Gupta2016}. The corresponding energy difference between $^1 A_1$ and $^1 E_1$ is 1042 nm. Thus, by applying this wavelength, one can change and monitor the $^1 E_1$ population. In an absorption-based measurement, however, an IR laser is used to probe the changes in the density of the metastable state, which can improve the signal contrast and collection efficiency and thus the sensitivity \cite{Jensen2014}. Jensen et al. achieved a magnetic field sensitivity of 2.5 nT$/\sqrt{Hz}$, and project a photon shot-noise-limited sensitivity of 70 pT$/\sqrt{Hz}$ for a few mW of infrared light at room temperature using IR-based cavity-enhanced magnetometry with NV centers in diamond, with the mirror spacing on the order of a few cm \cite{Jensen2014}. Furthermore, the transition between the singlet states $^1 E$ and $^1 A_1$ demonstrates a saturation intensity ($I_{\text{sat}}^{1042} \approx$ 0.5 W/$\mu$m$^2$) \cite{Dumeige_2013,Barry_2020}, which did not impose limitation in that work. However, saturation of this transition can result in constraints in particular for smaller cavities. Chatzidrosos et al. \cite{Chatzidrosos_2017} achieved a magnetic field sensitivity of 28 pT$/\sqrt{Hz}$ using a sensing volume of 390 $\mu$m $\times$ 4500 $\mu$m$^2$.

\par
Previously, for calculating magnetic sensitivity based on IR absorption, the spin density of the levels in NV centers within a cavity was determined using a rate equation, which considered only spin-conserving optical transitions \cite{Dumeige_2013}. In this work, we advance the methodology by calculating the spin density of the levels through quantum master equation calculations, providing a more comprehensive and precise understanding of the system's dynamics. To achieve this, we assume a model integrating diamond within a Fabry–Pérot cavity and explore the effects of cavity properties on sensitivity for a given spatial resolution. Our approach is in a reasonable agreement with  the experimental observation for larger cavities by Chatzidrosos et al. \cite{Chatzidrosos_2017}. We also use the differential evolution method \cite{Rocca2011} to find the optimized range of hyperparameters to achieve the best possible sensitivity. 

\par
This paper is organized as follows: In section II, we use ODMR studies to calculate IR absorption of NV center in diamond. Next, we calculated the transmitted and reflected light intensities using a Fabry-Pérot cavity model, including IR absorption, and quality factor corresponding to the intrinsic losses. Based on these findings and incorporating saturation limits and input power, we calculate the sensitivity of the NV center in the cavity. Section III discusses further aspects of this research. 

\section{Results}
\subsection{The spin dynamics}
In NV centers even when there is no external magnetic field present, the $m_s=0$ and $m_s= \pm 1$ states are non-degenerate due to the strong dipolar interaction between electronic spins, known as zero-field interaction or zero-field splitting. The spin Hamiltonian of the system in the eigenframe reads as follows:

\begin{align}
\hat{H} &= \sum_{i=x,y,z} D_i \hat{S}_i^2  + \mu_B g_e \hat{S}_z B_z = D \left( \hat{S}_z^2 - \frac{1}{3} S(S+1) \right) \nonumber \\
&\quad + E (\hat{S}_x^2 - \hat{S}_y^2) + \mu_B g_e \hat{S}_z B_z,
\label{eq:eq1}
\end{align}
where the axial component of the magnetic dipole–dipole interaction $D=\frac{3}{2}D_z$ and the transversal component $E=\frac{D_x-D_y}{2}$. $\mu_B$ and $g_e$ are the Bohr magneton and the electron gyromagnetic ratio, respectfully. $D$ describes the axial component of the magnetic dipole–dipole interaction, and $E$ the transversal component

We obtain the time evolution of the density matrix by using the Lindblad master equation:

\begin{align}
\frac{d\hat{\rho}}{dt}=\mathcal{L}\hat{\rho}=-i [H,\hat{\rho}]+\sum_{i=1}^{N} \gamma_i \big[L_i \hat{\rho} L_i^{\dagger}-\frac{1}{2} \{L_i^{\dagger} L_i, \hat{\rho}\} \big]
    \label{eq:eq2},
\end{align}
where \(L_i\) and \(\gamma_i\) are the jump operators and their corresponding rates. For further details on the master equation, see Appendix B, where Table \ref{tab:Table1} includes details about \(L_i\) and \(\gamma_i\).

\subsection{Absorption-based optically detected magnetic resonance}
In this work, we focus on the ODMR signal using an IR absorption-based approach. The ODMR signal is calculated by solving the master equation, given by Eq. \ref{eq:eq2}, for the NV$^-$ center under the influence of the IR laser, optical pump, and microwave (MW) fields. We account for dissipation, dephasing, coherent MW drive, and optical pumping, including the effects of both the IR and optical lasers, by incorporating their respective Lindblad superoperators (see Appendix B). The Hamiltonian for the NV$^-$ center is constructed by considering the ground, excited, and metastable states, depicted in Fig. \ref{fig:Figure1}, along with their corresponding electronic and spin energy levels. Additionally, we include the effects of the external magnetic field and coherent MW drive. The steady-state solution of the master equation is obtained in the rotating frame of the MW drive to determine the ODMR signal. The change in IR absorption is calculated by monitoring the population of the IR-absorbing metastable state. In this context, the ODMR signal is computed by measuring the difference in the population of the NV$^-$ center's metastable states with and without the MW drive, and subsequently calculating the IR absorption as a function of MW detuning.

\begin{figure}[ht!]
\centering
\includegraphics[width=3.2in]{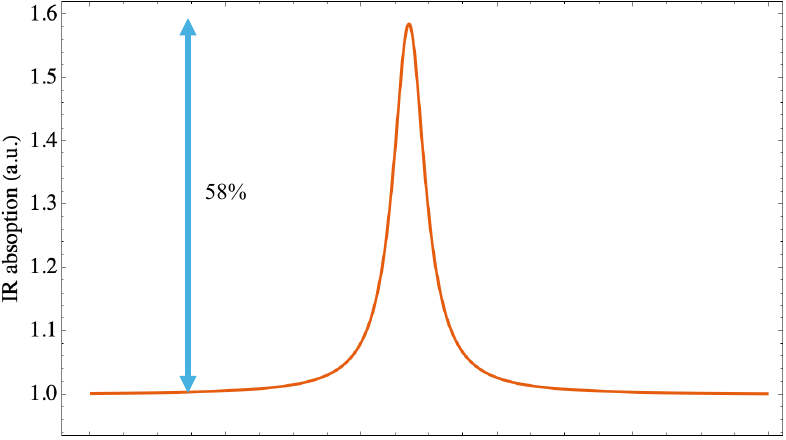}
\caption{The IR absorption ODMR signal for a single NV at zero static (bias) magnetic field.}
\label{fig:Figure2}
\end{figure}

The simulation parameters are typical of experimental setups and have been validated by reproducing fluorescence-based ODMR results (see Appendix B). The simulated absorption-based ODMR signal for a system with MW drive Rabi frequency $\Omega_R = 5$ MHz and $T_2^* = 500$ ns is presented in Fig. \ref{fig:Figure2}. The ODMR contrast for this setup is 58$\%$, representing the increase in absorbed IR photons relative to the far-detuned case. This approach enables a more realistic calculation of the steady-state density of states for the electronic structure of the NV$^-$ center, accounting for decoherence mechanisms. The results are then applied to the subsequent analysis of magnetic field sensitivity.

\subsection{Magnetometry}
\begin{figure*}
    \centering
 \includegraphics[width=0.7\textwidth]{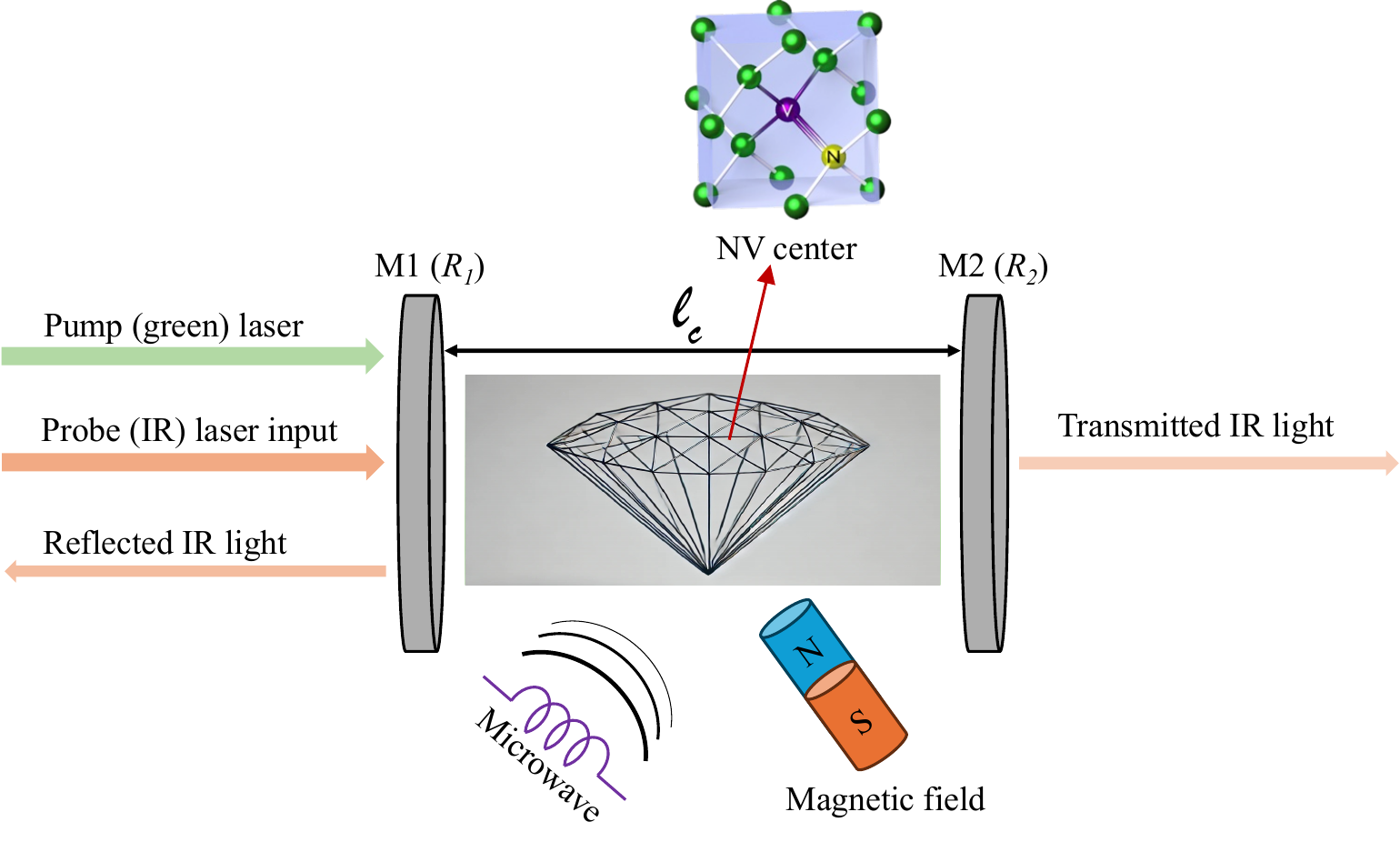}
    \caption{Schematic presentation of magnetometry based on NV center in diamond in Fabry-Pérot cavity model. In the model we use a green (532 nm) laser as the pump and an IR (1042 nm) laser as a probe. M$_1$ ($R_1$) and M$_2$ ($R_2$) are the mirrors with corresponding reflectivities R1 and R2. A magnetic field from a static magnet splits the NV spin sublevels that can be addressed by applying microwave signal. Note that our model makes no specific assumption about the interaction of the green pump laser with the cavity.}
        \label{fig:Figure3}
\end{figure*}

For the cavity, we consider a Fabry-Pérot model with two distinct mirrors, characterized by reflectivities \( R_1 \) and \( R_2 \), a cavity length \( l_c \), and light with a wavelength \( \lambda \), as shown schematically in Fig. \ref{fig:Figure3}. In this model, we also consider background loss, $a \equiv a(l_c,Q_i)=\frac{l_c}{\lambda Q_i}$, where $Q_i$ is the quality factor corresponding to the background loss. Of note, our model does not make any particular assumptions regarding the interaction between the green pump laser and the cavity.

For sensitivity calculation, our readout is the population difference between the singlet ground ($^1 E_1$) and excited ($^1 A_1$) states, $\Delta \rho$. We incorporated them into the reflected light intensity. Through this, we also included the saturation limit in terms of IR Rabi frequency, where the excited state population becomes comparable to the ground state population. This IR absorption depends on multiple factors such as NV center density ($n_{\text{NV}}$), and absorption cross-section ($\sigma_{\text{NV}}$). This can be denoted as optical depth $d=n_{\text{NV}} l_c \sigma_{\text{NV}} \Delta \rho$. The reflected light power intensity reads as follows:

\begin{align}
P_{ref}= \frac{P_{in} (\sqrt{R_1} - \sqrt{R_2} e^{-(a+d)})^2+4\sqrt{R_1 R_2} e^{-(a+d)} \sin^2{\phi}}{(1-\sqrt{R_1 R_2} e^{(a+d)})^2+4\sqrt{R_1 R_2} e^{-(a+d)} \sin^2{\phi}},
\label{eq:eq3}
\end{align} 
where $P_{in}$ and $\nu$ are the incoming light power intensity and the light frequency, respectively, and $\phi=2 \pi \nu l_c/c$ ($c=299792458$ m/s).  

The normalized reflected light, $P_{ref}$ in the presence of NV centers and IR absorption (for $R_1=0.9, R_2=0.99, Q_i=10^6, n_{NV}=10^{24}$ m$^{-3}$, the incoming light power $P_{in}=80$ mW, and $l_{c}=10^{-5} m$) is shown in Fig. \ref{fig:Figure4}.

The circulating light intensity in the cavity can be written as 
\begin{align} 
I_{cir}= \frac{I_{inc}(1-R_1)e^{-(a+d)}}{(1-\sqrt{R_1 R_2} e^{-(a+d)})^2+4\sqrt{R_1 R_2} e^{-(a+d)} \sin^2{\phi}}.
\label{eq:eq4}
\end{align}

\begin{figure}
    \centering
    \includegraphics[width=3.3in]{Figure4.pdf}
    \caption{The reflected light power, $P_{ref}$, vs. offset microwave frequency in the presence of NV centers with the incoming light power of $P_{in}=80$ mW. $\mathcal{C}$ is the contrast and $\Delta \nu$ is the full width at half the maximum of the magnetic resonances.}
    \label{fig:Figure4}
\end{figure}

The magnetic field sensitivity of the cavity with NV centers, $\delta B$, can be estimated based on the minimum detectable magnetic field during an acquisition time taking into account photon shot noise,  which depends on the contrast, $\mathcal{C}$,  and the full width at half maximum, $\Delta \nu$, of the magnetic resonances \cite{Taylor2008,Jensen2014}, as shown in Fig. \ref{fig:Figure4}:
\begin{align}
\delta B=\frac{2 \pi \Delta \nu}{\gamma \mathcal{C}} \sqrt{\frac{h \nu }{ P }},
\label{eq:eq5}
\end{align}
where $h$, and $\nu$, $\gamma=1.761$ $\times$ $10^{11}$ rad s$^{-1}$ T$^{-1}$ and $P$ are, respectively, the Planck constant, photon frequency (IR light 1042 nm), the gyromagnetic ratio and the reflected power.
\par
We also considered the spin shot noise limit \cite{Dumeige_2013}, 
\begin{align}
\delta B_s=\frac{2}{\gamma \sqrt{n_{NV} l_c \sigma_m T^{*}_{2} t_m}},
\label{eq:eq6}
\end{align}
where $\sigma_m$ and $t_m$ are, respectively, the mode size and measurement time, and we ensured that it was at most 20\% of $\delta B$, where $T^{*}_{2}=2/\Delta \nu$. Notably, in the presence of an external magnetic field, the spin \(\pm 1\) sub-levels will split due to the Zeeman interaction, resulting in two resonance peaks in the ODMR signal. Additionally, considering the four different orientations of NV centers in the diamond crystal, these peaks will further split into multiple peaks. When the external magnetic field is aligned with one of these orientations, four peaks will be observed: two from the aligned NVs and two from the three other orientations, as shown in Fig. \ref{fig:Figure4}.

\par
In their research on IR-based magnetometry, Chatzidrosos et al. utilized a cavity composed of two mirrors with reflectivities of $R_1=98.5\%$ and $R_2=99.2\%$ at 5 mm from each other, a 0.39 mm diamond length featuring a high NV center density of $28 \times 10^{23}$ m$^{-3}$, and an input power of $P_{in}=80$ mW \cite{Chatzidrosos_2017}. This setup enabled them to achieve a magnetic field sensitivity of 28 pT$/\sqrt{\text{Hz}}$ with a spin noise of $\sim$ 0.46 pT$/\sqrt{\text{Hz}}$. To validate our model, we aim to reproduce their achieved magnetic field sensitivity based on the used setup. As our model treats the cavity length as the diamond length, we explored two cases. In the first case, we considered that the cavity's length is the same as their diamond, where we used $28 \times 10^{23}$ m$^{-3}$ as the NV center density. For this case, our model resulted in 5 pT$/\sqrt{\text{Hz}}$. For the second case, we used a cavity length similar to theirs but set the NV center density to an effective value of $2.18\times$10$^{23}$ m$^{-3}$ in order to match the optical depth per pass in their experiment. Our model, for this set, gave a sensitivity of 2.5 pT$/\sqrt{\text{Hz}}$. Based on this theoretical model, the spin noise was $\sim$ 0.25 pT$/\sqrt{\text{Hz}}$. Our model predicts somewhat better sensitivity than is observed experimentally, which is likely due to to a combination of the monolithic character of the cavity in our model, as well as additional technical noise sources in the experiment in addition to photon shot noise and spin shot noise. 

\par
Using Eq. \ref{eq:eq5}, to achieve optimal values, we explored different values for the mirrors' reflectivities, intrinsic loss, NV center density, and cavity length and cross-section. Optimizing hyperparameters for a cavity-based sensing system presents a complex challenge due to the inherently non-linear nature of the parameters involved. To address this complexity, we employed the Differential Evolution algorithm \cite{Rocca2011}. The optimization of parameter values for sensing performance was carried out considering the following parameter bounds: \(R_1\) and \(R_2 \in [0.9, 0.999999]\), \(\ell_c \in [10^{-7} \, \text{m}, 10^{-2} \, \text{m}]\), \(\sigma_m \in [10^{-14} \, \text{m}^2, 10^{-4} \, \text{m}^2]\), \(P_\text{in} \in [0.01 \, \text{W}, 1 \, \text{W}]\), and \(n_\text{NV} \in [10^{17}, 10^{24}]\). In all cases, we set \(Q_i = 10^6\), with the optimization targeting optimal sensing performance.

Additionally, we accounted for constraints imposed by spin noise and the saturation limits of the system, ensuring that the optimization process remained grounded in the physical realities of the sensing mechanism. Specifically, the intra-cavity intensity is constrained by the saturation intensity of 0.5 W/$\mu$m$^2$ \cite{Jensen2014}. In this work, the spin projection noise was calculated for each configuration, and we ensured that its contribution remained below 20\% of the photon shot noise, unless stated otherwise. This threshold was used as a guiding limit in our optimization to ensure the system's noise performance was dominated by photon shot noise, while still accounting for the effects of spin projection noise.

\par

\begin{figure}[t]
\centering
    \includegraphics[width=3in]{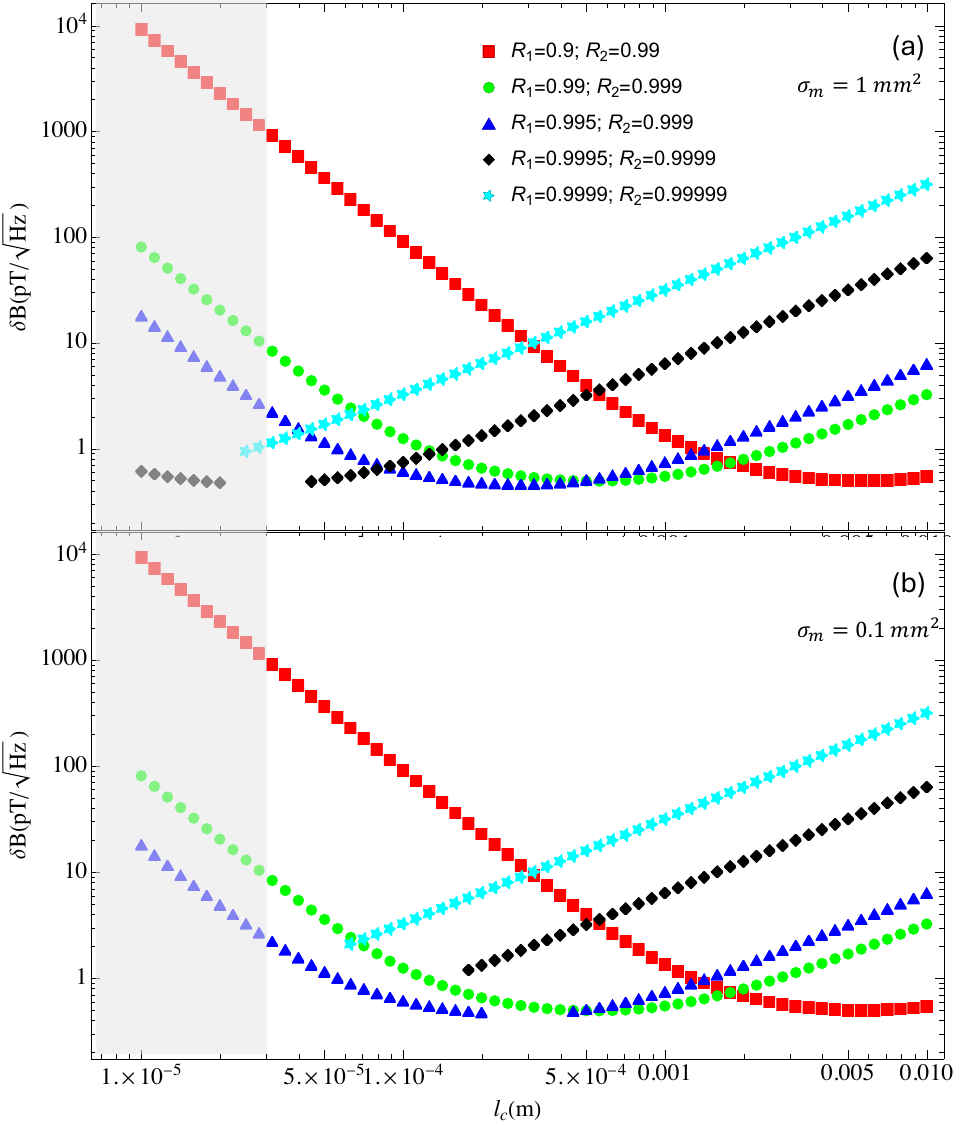}
    \caption{Magnetic Field Sensitivity vs Cavity Length. The sensitivity vs the cavity length is calculated for different mirror reflectivities (see legend) for input power of 1 W, quality factor of 10$^6$, NV density of 10$^{24}$ m$^{-3}$, and mode cross-section of 1 mm$^2$ (a) and 0.1 mm$^2$ (b). Here, the intra-cavity intensity is below the saturation limit of 0.5 W/$\mu$m$^2$. The spin shot noise limit is below 20\% of the photon shot noise. The gray area indicated lengths smaller than $\sim$ 30 $\mu$m (10 $\mu$m diameter for micro-cavity).}
    \label{fig:Figure5}
\end{figure}


\par
Fig. \ref{fig:Figure5} shows the sensitivity as a function of the cavity length for different mirror reflectivities and different cavity dimensions, where the spin noise and intra-cavity saturations limits were imposed. Notably, Fig. \ref{fig:Figure5} demonstrates that the sub-pT sensitivity is achievable over a wide range of cavity lengths.

\par
\begin{figure*}
    \centering
 \includegraphics[width=0.9\textwidth]{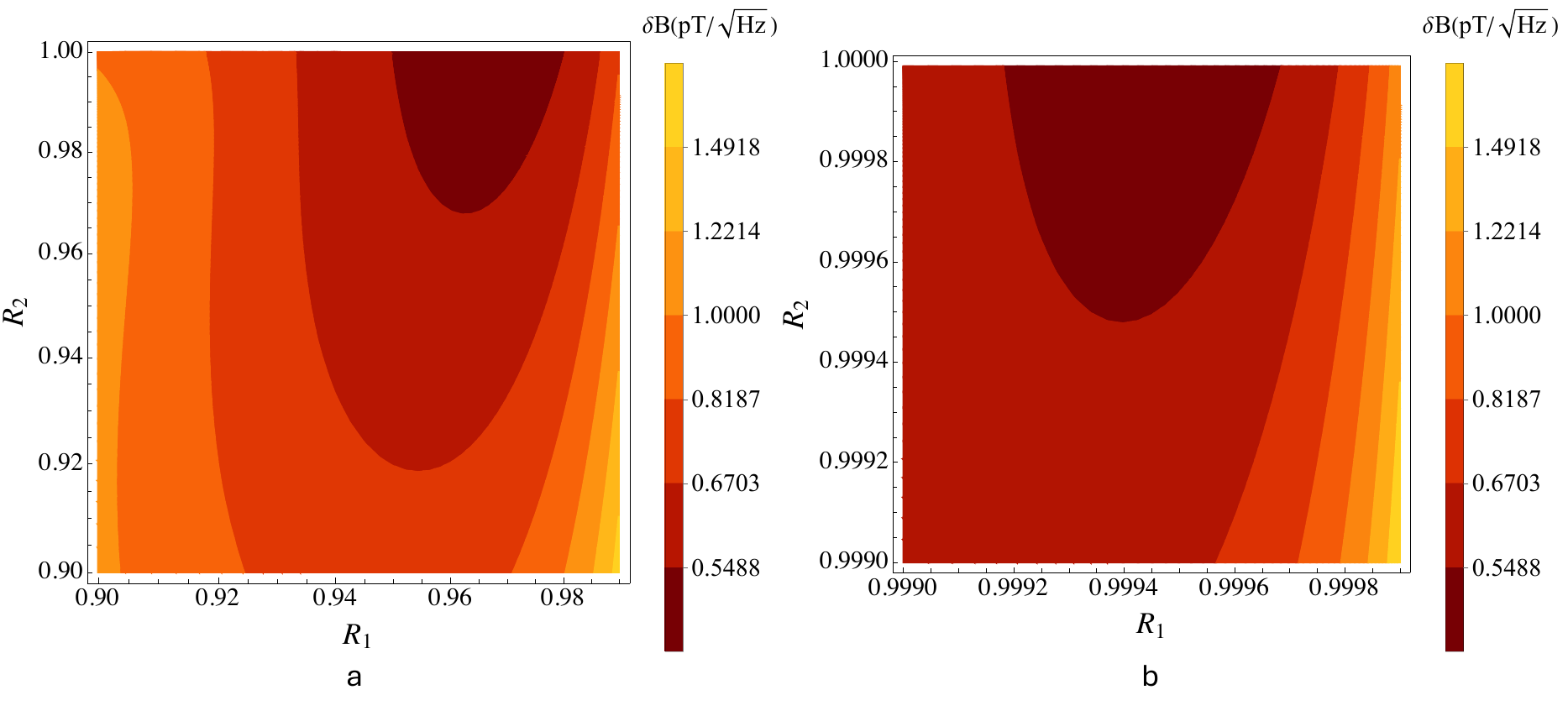}
    \caption{Magnetic-field sensitivity as a function of the two mirror reflectivities for different dimensions, (a) with the cavity length of 3 mm and mode cross-section of 1 cm$^{2}$ and (b) with the cavity length of 50 $\mu$m and mode cross-section of 1 mm$^{2}$. Both cases can achieve sub-pico tesla sensitivity. For such sensitivity having smaller dimensions requires higher mirror reflectivities. In both cases, the intra-cavity intensity is below the saturation limit of 0.5 W/$\mu$m$^2$ and the spin shot noise limit is below 20\% of the photon shot noise. The cavity quality factor, NV density in diamond, and input power are 10$^6$, 10$^{24}$ m$^{-3}$, and 1 W, respectively.}
        \label{fig:Figure6}
\end{figure*}

For experimental implementation purposes, by fixing the quality factor, input power, cavity length, and NV center density, we explored the impact of mirrors' reflectivities, as shown in Fig. \ref{fig:Figure6}. For a length of 1 cm, the sensitivity can reach sub-pico tesla. For a smaller cavity, even higher reflectivities are required for achieving the same order of sensitivity (Fig. \ref{fig:Figure6}b). Further hyperparameter exploration can be found in the Appendix. 

\begin{figure*}
    \centering
 \includegraphics[width=0.9\textwidth]{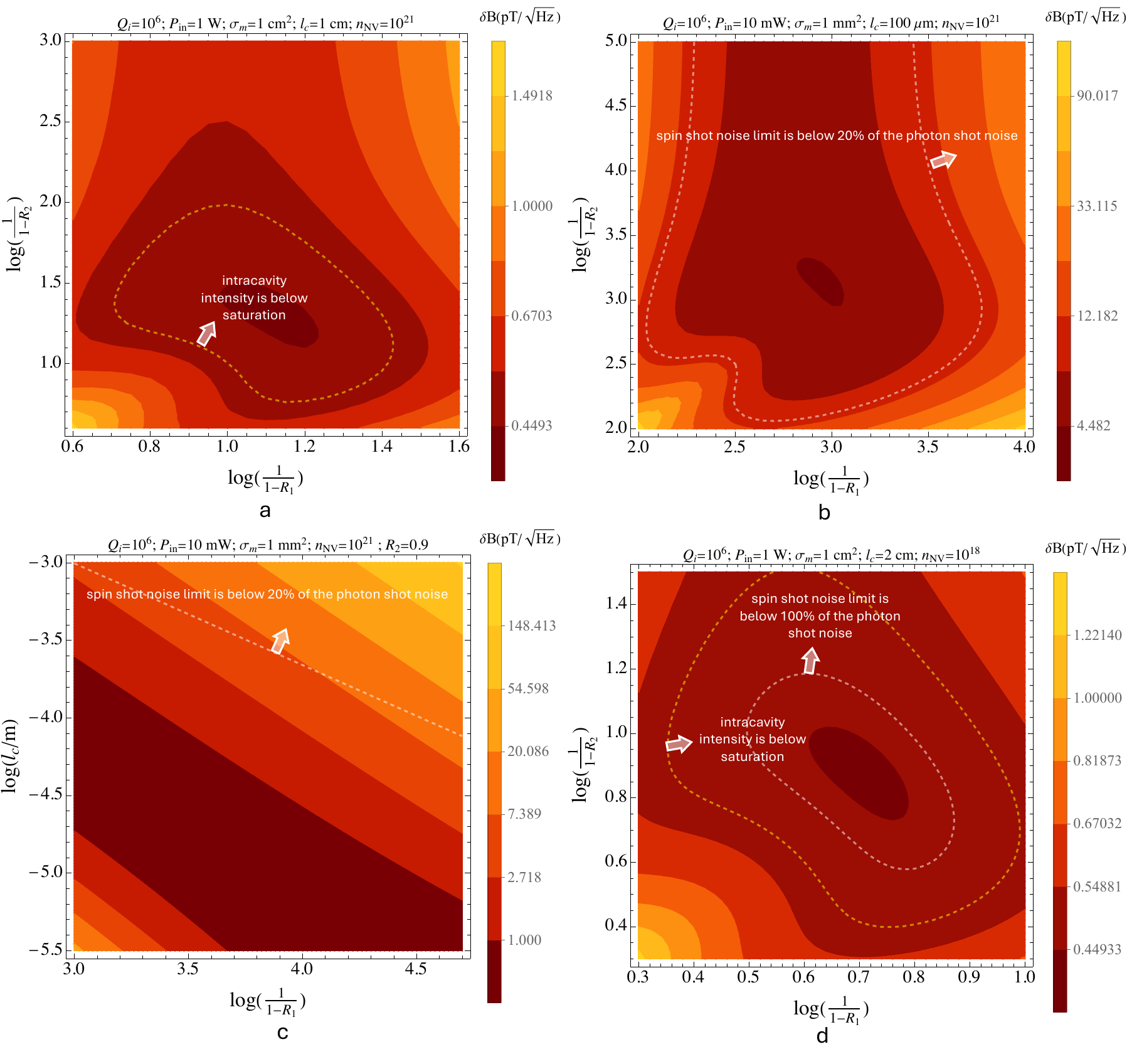}
    \caption{Sensitivity for lower NV densities. Magnetic-field sensitivity on the mirrors' reflectivities (a) and (b) and on mirror-1 reflectivity and cavity length (c) for NV density of 10$^{21}$ m$^{-3}$. The dashed lines indicate the parameter region where the spin shot noise limit is below 20\% of the photon shot noise. (d) Magnetic-field sensitivity on the mirrors' reflectivities for the cavity length for NV density in dimond of 10$^{18}$ m$^{-3}$. The inner dashed line indicated the region where the spin noise shot is below 100\% of the photon shot noise. The Larger dashed line indicated the regio where the intra-cavity light intensity is below the saturation limit of 0.5 W/$\mu$m$^2$. In all cases, the absence of the dashed lines for both constrains implies that the condition are satisfied.}
        \label{fig:Figure7}
\end{figure*}

We also explored scenarios where the NV center density in diamond were 10$^{21}$ m$^{-3}$ and 10$^{18}$ m$^{-3}$, as shown in Figs. \ref{fig:Figure7}(a-c) and \ref{fig:Figure7}(d),respectively. Having lower NV center densities results in higher spin noise. For 10$^{21}$ m$^{-3}$, to achieve sub-pico tesla one needs larger cavity lengths. To get the same order of sensitivity, using 10$^{18}$ m$^{-3}$ requires not only larger cavity lengths, but also we had to allow the spin noise to become equal to the photon shot noise.

\par

\section{Discussion}

In this study, we explored the potential of using diamond-based Fabry-Pérot cavities for enhancing the performance of magnetometry using NV centers based on IR absorption of the singlet states. Using the Lindblad master equation, we calculated the population of singlet states, including the saturation limits. The master equation gives us the singlet population, which we then feed into the calculation of the reflectivity, from which we derive the magnetic field sensitivity. Using the differential evolution algorithm, we were able to explore the hyperparameter space, including mirror reflectivities, cavity length, cross-section mode size, quality factor, input power, and NV center density. For this optimization search, we also imposed the constraints due to the saturation and spin noise limitations. This analysis shows that even for smaller cavities, one can achieve high sensitivities of sub-pico tesla.
\par
Of note, in our calculations, we specifically considered the length of the Fabry-Pérot cavity to be equivalent to the length of the diamond, a scenario typical for NV centers in diamond within cavities \cite{Shandilya2021,Mitchell2016}. This is a valid assumption for monolithic cavities. However, for free-space Fabry-Pérot cavities, where the cavity length may not necessarily match the medium length there can be additional reflections at the interfaces (e.g., diamond), further care must be exercised.  
\par
Our model allows for exciting extensions, particularly in the experimental exploration of microcavities, which could enhance NV center-based magnetometry by achieving higher sensitivities and spatial resolutions. Highly sensitive NV magnetometers have promising applications in medical diagnostics \cite{Kuwahata2020}, such as magnetocardiography (MCG) and magnetoencephalography (MEG), which require sensitivities of order 1 pT \cite{Fenici2005} and 1 fT \cite{Hmlinen1993}, respectively. While current state-of-the-art SQUIDs and optically pumped atomic magnetometers achieve these sensitivities, their spatial resolution is limited to mm-scale. In contrast, microcavity based NV magnetometers, operating at room temperature, could offer a few micron-scale spatial resolution, enabling non-invasive, high-resolution imaging of brain and heart activity.

Magnetometers based on on chip microcavity design provide several key advantages over the standard Fabry-Pérot cavity design. Their smaller size renders the sensor head to have a smaller footprint in \textit{in vivo} applications thus simplifying the endoscopic techniques. Such a miniaturized sensor obviously allows for reduced stand-off distance between the sensor and the specimen, a key factor that determines the sensitivity. For instance, extracellular sensing of magnetic fields \cite{Barry2016} generated by neuronal action potentials using these sensors would highly simplify the existing biomedical tools for such applications, potentially allowing label-free measurements at the single neuron level with minimal photodamage to the specimen.

Besides offering scalability, the ability to fabricate near-identical microdisk cavities in a grid-like fashion on the same diamond chip enables realizing gradiometric sensing schemes that achieve high throughput sensing with $\mu$m spatial resolution on spatially inhomogeneous samples in condensed matter and biomedical areas.

Beyond biomedical applications, cavity-based NV magnetometers enhanced sensitivities hold potential in corrosion detection \cite{Juzelinas1999,Bardin2015,Komary2023}, materials characterization of static magnetic properties \cite{Romalis2011}, altermagnetism in condensed matter \cite{mejkal2022}, quantum gravimeters and gradiometers for navigation in remote and inaccessible locations \cite{Witze2019,Fu2020,Wang2023} and for monitoring underground water levels and volcanic eruptions \cite{AntoniMicollier2022}.

Additionally, the four distinct orientations of the NV axis in diamond allow for vector magnetic field imaging, offering a significant advantage over other magnetometers, which typically measure only the magnitude of the magnetic field. This capability enables more comprehensive and precise mapping of magnetic field directions and strengths in various applications \cite{Clevenson2018,Yahata2019,verctor}.

\section*{Acknowledgment}
The authors thank Erika Janitz and Faezeh Kimiaee Asadi for helpful discussions. This work was supported by an NSERC/Alberta Innovates Advance Grant.

\appendix

\section{Sensitivity dependence on various combinations of parameter}
\begin{figure*}[ht!]
    \centering
 \includegraphics[width=0.9\textwidth]{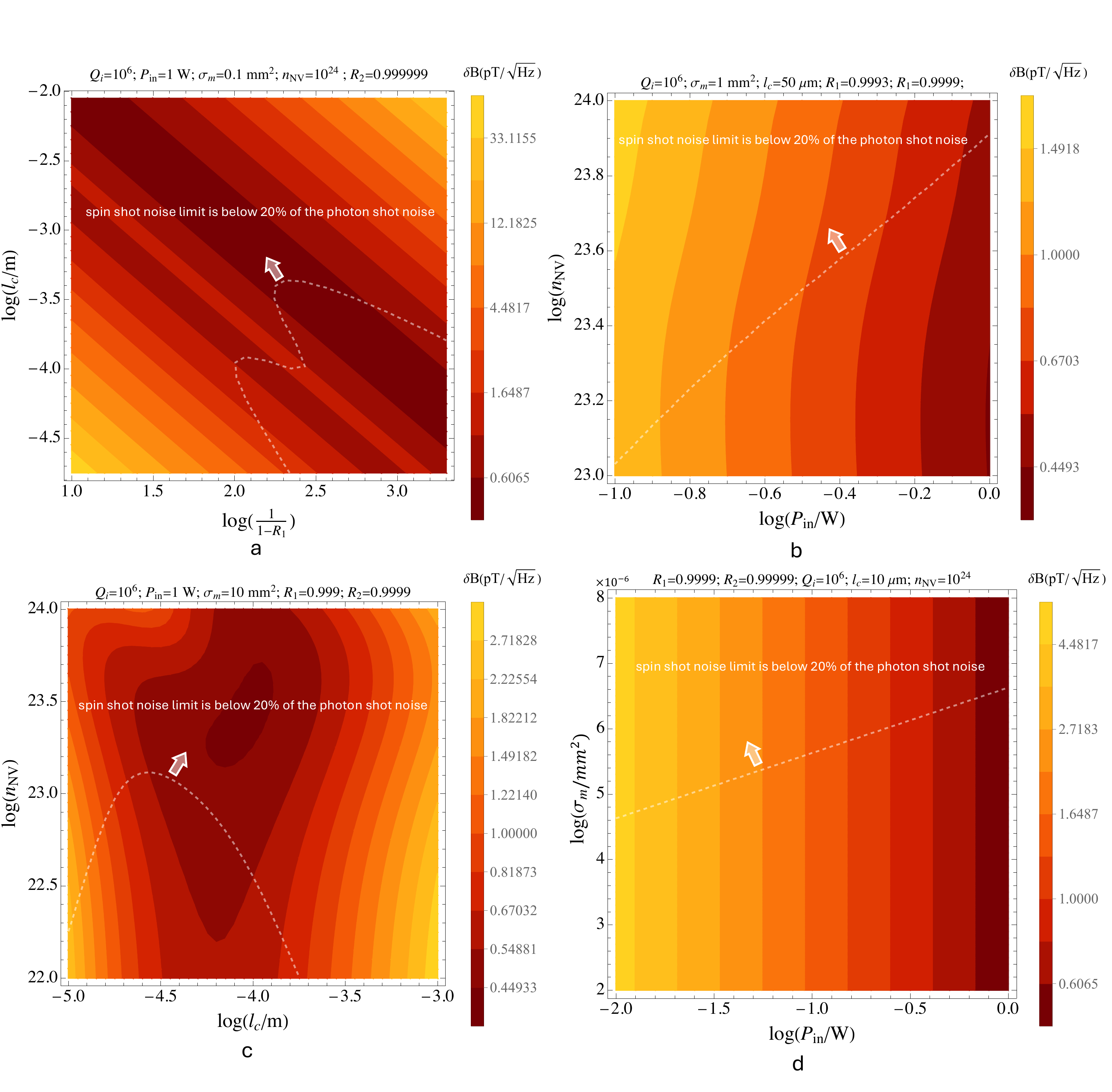}
    \caption{Sensitivity dependence on various combinations of parameters. The sensitivity (a) on mirror-1 reflectivity and cavity length, (b) on the NV density in diamond and input power, (c) on NV center density in diamond and cavity length, and (d) on mode cross-section and input power. In all case the sensitivity can reach to sub-pico tesla. The dashed lines indicate the parameter region where the spin shot noise limit is below 20\% of the photon shot noise. In all cases, the absence of the dashed lines for both constraints implies that the conditions are satisfied.}
        \label{fig:Figure8}
\end{figure*}

Fixing mirrors' reflectivities, input power, quality factor, and mode cross-section size, we searched the cavity length and NV center density in diamond, considering the limitations due to the noise and intensity saturation in the cavity, the sensitivity can reach to a sub pT/\(\sqrt{\text{Hz}}\), as shown in Fig. \ref{fig:Figure8}. We also looked at other cases where for having higher sensitivity and smaller cavities higher mirror reflectivities are required. Further, we investigated extreme cases where the input power was considered as a variable and cavity size of 10 $\mu$m and mode cross-section of 100 $\mu$m$^2$. Due to the saturation and noise limits resulted in sensitivities order of a sub pT/\(\sqrt{\text{Hz}}\), as shown in Fig. \ref{fig:Figure8}c. 

\section{Master equation}
\begin{figure}
\centering
\includegraphics[width=2.5in]{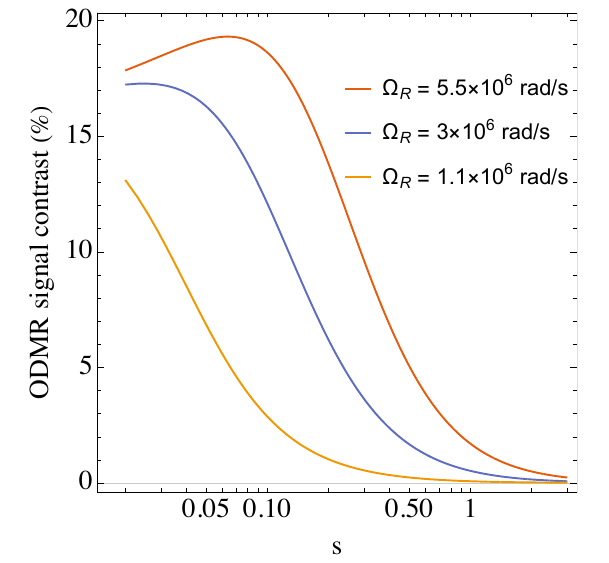}
\caption{ODMR contrast as a function of s saturation parameter $s=P_{in}/P_{sat}$, where $P_{in}$ and $P_{sat}$ are the optical pumping power and the saturation power of the transition, respectively and $P_{sat}$=25 MHz, based on the setup in Dréau et al. \cite{Drau2011}.}
\label{fig:Figure9}
\end{figure}

To simulate the ODMR signal, we initially compared a fluorescence-based signal, rather than an IR absorption-based one, for a single NV center with experimental parameters. This comparison was conducted to evaluate the accuracy of our theoretical model in reproducing the observed results in \cite{Drau2011}. Fig. \ref{fig:Figure9} illustrates the fluorescence-based ODMR contrast as a function of the saturation parameter \(s = P_{\text{in}}/P_{\text{sat}}\), where \(P_{\text{in}}\) represents the optical pumping power and \(P_{\text{sat}}\) denotes the saturation power of the transition. Here, \(P_{\text{sat}} = 25\) MHz, based on the setup in Dréau et al. This comparison serves exclusively to validate the master equation solver and the parameters employed within it. The only major change in the master equation analysis compared to the IR absorption simulation is detecting optical photoluminescence instead of IR absorption. Furthermore, the jump operators used in the master equation analysis are summarized in Table \ref{tab:Table1}.

\begin{table}[h]
    \centering
    \begin{tabular}{|c|c|c|}
    \hline
    \textbf{Operator} & \textbf{Matrix Element} & \textbf{ Rate} \\ \hline
    \multicolumn{3}{|c|}{\textbf{Optical Transitions}} \\ \hline
    $L_1$ & $|1\rangle \langle 4|$ & $\gamma_1 = k_{41}$ \\
    $L_2$ & $|2\rangle \langle 5|$ & $\gamma_2 = k_{52}$ \\
    $L_3$ & $|3\rangle \langle 6|$ & $\gamma_3 = k_{63}$ \\
    $L_4$ & $|4\rangle \langle 1|$ & $\gamma_4 = k_{41} + W_p$ \\
    $L_5$ & $|5\rangle \langle 2|$ & $\gamma_5 = k_{52} + W_p$ \\
    $L_6$ & $|6\rangle \langle 3|$ & $\gamma_6 = k_{63} + W_p$ \\
    $L_7$ & $|7\rangle \langle 8|$ & $\gamma_7 = k_{87}$ \\
    $L_8$ & $|8\rangle \langle 7|$ & $\gamma_8 = k_{87}$ \\ \hline
    \multicolumn{3}{|c|}{\textbf{Spin Transitions}} \\ \hline
    $L_9$ & $|1\rangle \langle 2|$ & $\gamma_9 = \gamma_{gMW}$ \\
    $L_{10}$ & $|2\rangle \langle 1|$ & $\gamma_{10} = \gamma_{gMW}$ \\
    $L_{11}$ & $|1\rangle \langle 3|$ & $\gamma_{11} = \gamma_{gMW}$ \\
    $L_{12}$ & $|3\rangle \langle 1|$ & $\gamma_{12} = \gamma_{gMW}$ \\
    $L_{13}$ & $|4\rangle \langle 5|$ & $\gamma_{13} = \gamma_{eMW}$ \\
    $L_{14}$ & $|5\rangle \langle 4|$ & $\gamma_{14} = \gamma_{eMW}$ \\
    $L_{15}$ & $|4\rangle \langle 6|$ & $\gamma_{15} = \gamma_{eMW}$ \\
    $L_{16}$ & $|6\rangle \langle 4|$ & $\gamma_{16} = \gamma_{eMW}$ \\ \hline
    \multicolumn{3}{|c|}{\textbf{Dephasing}} \\ \hline
    $L_{17}$ & $|2\rangle \langle 2|$ & $\gamma_{17} = 2\gamma_{\Phi}$ \\
    $L_{18}$ & $|3\rangle \langle 3|$ & $\gamma_{18} = 2\gamma_{\Phi}$ \\
    $L_{19}$ & $|5\rangle \langle 5|$ & $\gamma_{19} = 2\gamma_{\Phi}$ \\
    $L_{20}$ & $|6\rangle \langle 6|$ & $\gamma_{20} = 2\gamma_{\Phi}$ \\ \hline
    \multicolumn{3}{|c|}{\textbf{Metastable States}} \\ \hline
    $L_{21}$ & $|8\rangle \langle 4|$ & $\gamma_{21} = k_{48}$ \\
    $L_{22}$ & $|8\rangle \langle 5|$ & $\gamma_{22} = k_{58}$ \\
    $L_{23}$ & $|8\rangle \langle 6|$ & $\gamma_{23} = k_{68}$ \\
    $L_{24}$ & $|7\rangle \langle 1|$ & $\gamma_{24} = k_{71}$ \\
    $L_{25}$ & $|7\rangle \langle 2|$ & $\gamma_{25} = k_{72}$ \\
    $L_{26}$ & $|7\rangle \langle 3|$ & $\gamma_{26} = k_{73}$ \\ \hline
    \end{tabular}
    \caption{Summary of jump operators. $k_{41}=k_{52}=k_{63}=66$ MHz, $k_{87}=1$ GHz, $k_{48}=7.9$ MHz, $k_{58}=k_{68}=53$ MHz, $k_{72}=k_{73}=0.7$ MHz, $k_{71}=1$ MHz, $W_p=1$ MHz, $\gamma_{gMW}=1$ kHz, and $\gamma_{\phi}=1$ MHz \cite{Dumeige_2013}. }
    \label{tab:Table1}
\end{table}

\bibliography{Ref}

\end{document}